\documentclass[12pt,a4paper,dvips]{article}
\usepackage{a4p}
\usepackage{cite,mcite}
\usepackage{graphicx}
\usepackage{physics }
\usepackage{l3_title,ifthen}
\usepackage{pennames}
\usepackage{hhline}
\usepackage{epsfig}
\journalname{Phys. Lett. B}
\preprint{2002-065}
\date{August 2, 2002}
\newlength{\capindent}
\newlength{\capwidth}
\newlength{\figwidth}
\newcommand{\icaption}[2][!*!,!]{\hspace*{\capindent}%
  \begin{minipage}{\capwidth}
    \ifthenelse{\equal{#1}{!*!,!}}%
      {\caption{#2}}%
      {\caption[#1]{#2}}
  \end{minipage}}

\def\BR{\mbox{BR}}
\def\CL95{95~\%~\mbox{C.L.}}
\def\LEP2{LEP~II}
\def\LEP{LEP }
\def\epem{\ensuremath{\mathrm{e^+e^-}}}
\newcommand{\Pup}{\ensuremath{\mathrm{u}}}

\newcommand{\Pcharm}{\ensuremath{\mathrm{c}}}

\newcommand{\Pbottom}{\ensuremath{\mathrm{b}}}
\newcommand{\Ptop}{\ensuremath{\mathrm{t}}}
\newcommand{\Paup}{\ensuremath{\bar{\mathrm{u}}}}

\newcommand{\Pacharm}{\ensuremath{\bar{\mathrm{c}}}}

\newcommand{\PZ}{\ensuremath{\mathrm{Z}}}
\def\Kg{\kappa_{\gamma}}
\def\Kz{\kappa_{\rm Z}}

\def\eetc{\Pep\Pem\rightarrow\Ptop\Pacharm}
\def\eetu{\Pep\Pem\rightarrow\Ptop\Paup}

\def\backqq{\Pep\Pem\rightarrow\Pq\Paq(\gamma)}
\def\backww{\Pep\Pem\rightarrow\PWp\PWm}
\def\backtt{\Pep\Pem\rightarrow\Pgt^+\Pgt^-}
\def\backzz{\Pep\Pem\rightarrow\PZ\PZ}
\begin{document}
\begin{titlepage}
\title{Search for Single Top Production at LEP}
\author{The L3 Collaboration}
\begin{abstract}
Single top production in \epem annihilations is searched for in data collected
by the L3 detector at centre-of-mass energies from 189 to 209~GeV, 
corresponding to a total
integrated luminosity of 634~pb$^{-1}$. 
Investigating hadronic and semileptonic
top decays, no evidence of single top 
production at \LEP is obtained and upper limits on the single 
top cross section as a function of the centre-of-mass energy are derived.
Limits on possible anomalous couplings,
as well as on the scale of contact interactions 
responsible for single top production are determined.
\end{abstract}
\submitted
\end{titlepage}
\section*{Introduction}
\label{introduction}
One of the fundamental features of the Standard Model is that  neutral 
currents are flavour diagonal, therefore any flavour changing neutral current
(FCNC)
process may occur only at second or higher orders through loops. 
The FCNC interactions of the top quark offer an ideal place to search for 
new physics and can be studied  either in $\Ptop\rightarrow\rm{qV}$ 
($\rm{q}=\Pup,\Pcharm$; $\rm{V}=\gamma,\PZ$) decays at the Tevatron or in 
single top production at LEP. In the Standard Model, the predicted rates 
of such processes are very small~\cite{Eilam}, but new physics 
beyond the Standard Model could lead to 
a significant increase. Enhancements of some orders of magnitude are
predicted in two-Higgs-doublet models and supersymmetric 
models~\cite{Li}. 
Flavour changing multiple-Higgs-doublet 
models further enhance the rates
up to $10^{-5}$~\cite{Cheng}, and models with FCNC coupled
singlet quarks, compositeness or dynamical
electroweak symmetry 
breaking~\cite{Barger} 
can yield a branching ratio of about  $10^{-2}$, in the reach of present 
colliders. The same considerations apply for the 
predicted single top production 
rates~\cite{Boos}.
Experimental upper limits at the 95\% confidence level
on the FCNC branching fractions of the top quark were set by the CDF 
Collaboration as
$\BR(\Ptop\rightarrow\Pcharm(\Pup)\gamma) <  3.2\%$ and 
$\BR(\Ptop\rightarrow\Pcharm(\Pup)\PZ) < 33\%$~\cite{Abe:1998fz}.
Other studies of single top production were carried out at 
LEP~\cite{singletopaleph}.
\par
In this Letter, we consider two theoretical models for single top production
$\eetc$~\footnote{Charge conjugate is assumed throughout this Letter.}:
an interpretation involving [$\Ptop\Pacharm\Pep\Pem$] contact interactions
and an approach which describes
the process in terms of anomalous couplings.
\subsection*{Single Top Production through Contact Interactions}
\begin{figure}
\begin{eqnarray*}
  \epsfig{file=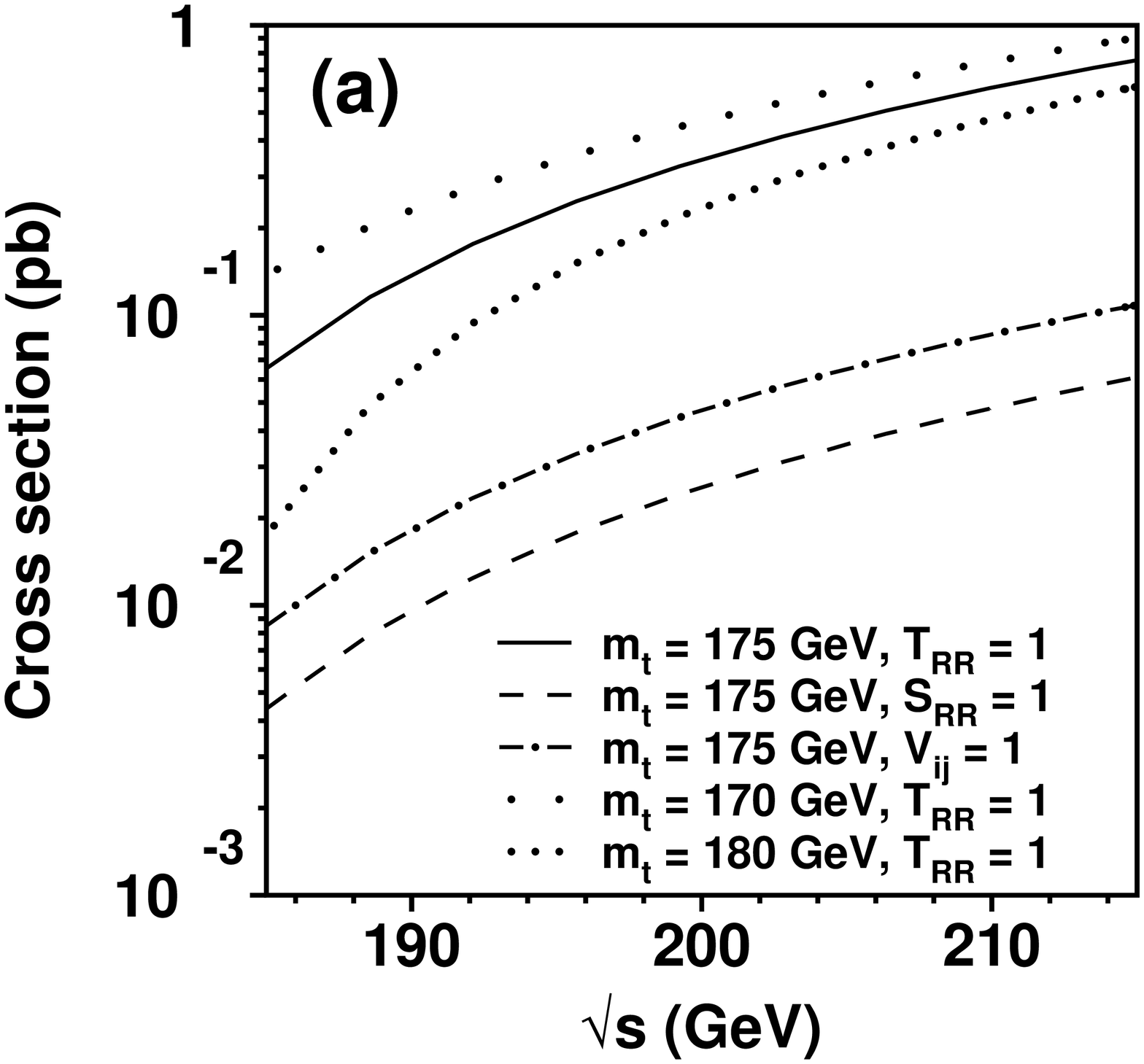,width=7.9cm} &
  \epsfig{file=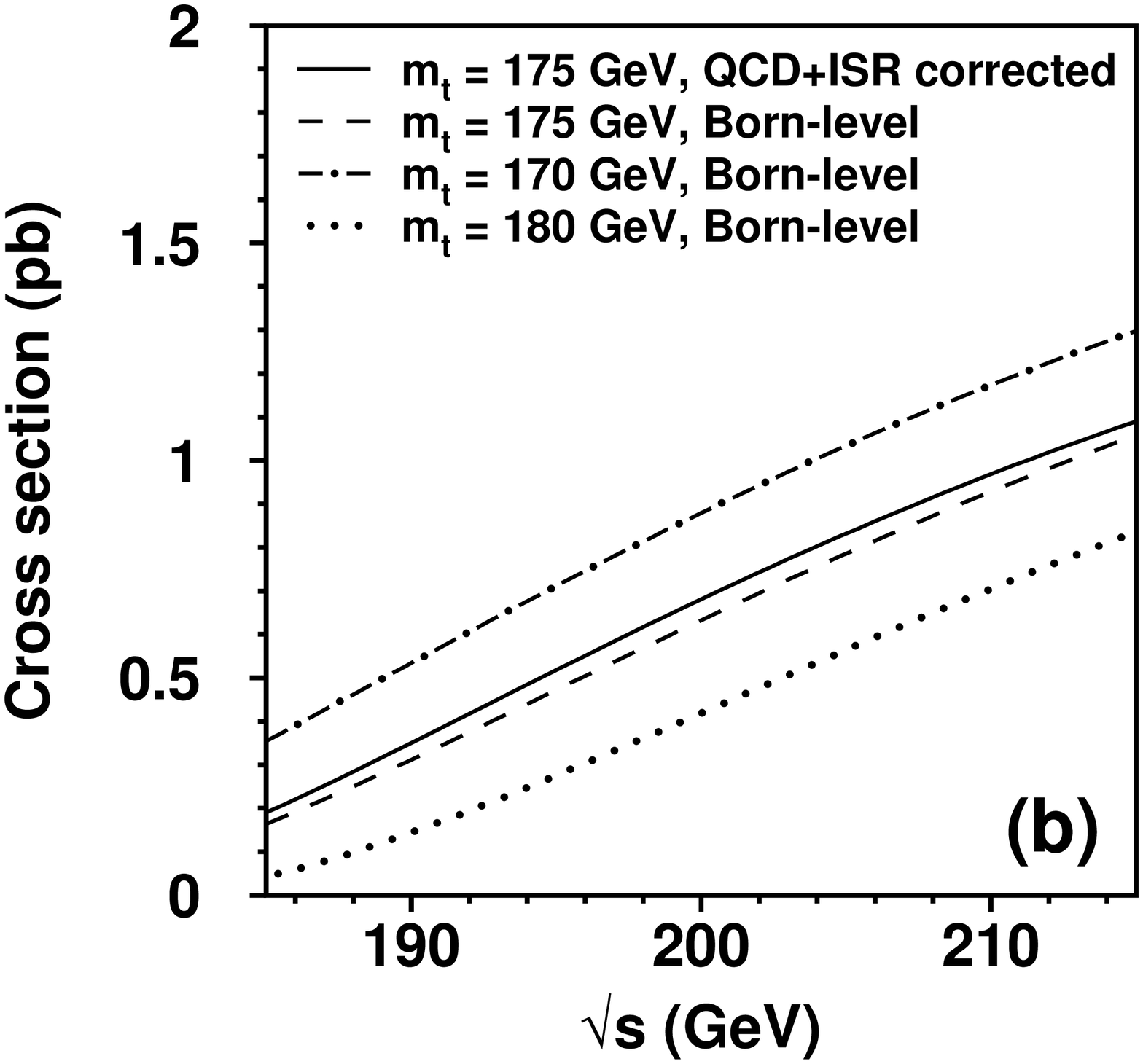,width=7.9cm} \\
\end{eqnarray*}
  \icaption{\rm
   Theoretical total cross section for single top production as a function of the
   centre-of-mass energy for three different top quark mass values
   for (a) [$\Ptop\Pacharm\Pep\Pem$] contact interactions with different
   assumptions on the Lorentz structure and an energy scale parameter
   $\Lambda =$ 1 TeV and 
   (b) the model described in Equation~(\ref{eq:cross_section}) where
   CDF limits on the values for the anomalous constants are assumed. 
    \label{fig:cross_sections}
    }
\end{figure}

An effective flavour changing vertex can be parameterized
through [$\Ptop\Pacharm\Pep\Pem$] contact interactions~\cite{Bar-Shalom:1999iy}.
The total unpolarised cross sections for $\eetc$ and charge conjugate can be 
written as:
\begin{equation}
\label{eq:four}
\sigma = \mathcal{C}\left[ 
  (3+\beta)(V_{LL}^{2} + V_{RR}^{2} + V_{RL}^{2} + V_{LR}^{2}) + 
  \frac{3}{2}(1+\beta)S_{RR}^{2} +
  8(3-\beta)T_{RR}^{2} 
\right]
\end{equation}
where $V_{ij}$, $S_{ij}$, $T_{ij}$ represent the coupling constants of vector,
 scalar and tensor fields, respectively (L and R stand for left-handed and
right-handed fields), 
$\mathcal{C}={s}/{\Lambda^{4}}\times{\beta^{2}}/{4\pi(1+\beta)^{3}}$,
$\beta=({s-m_{\Ptop}^{2}})/({s+m_{\Ptop}^{2}})$ and $\Lambda$ is the energy 
scale parameter for the process.
The predicted single top production cross sections are shown in 
Figure~\ref{fig:cross_sections}a
for three different assumptions on the Lorentz structure of the 
[$\Ptop\Pacharm\Pep\Pem$] vertex.
In this Letter, we give lower limits on the energy scale $\Lambda$. No directly
comparable previous limits exist.
\subsection*{Anomalous Couplings}
The Born-level cross section for the single top production process
$\eetc$ in the presence of
anomalous couplings $\Ptop\rm{V}\Pacharm$, with $\rm{V}=\gamma,\PZ$
~\cite{Obraztsov:1998if,Han}, is given by:
\begin{equation} 
\label{eq:cross_section}
\sigma = 
      \frac{\pi \alpha^2}{s}\biggl(1-\mathcal{M}_\Ptop\biggr)^2 
      \biggl[ \frac{\Kg^2 e_\Pq^2}{\mathcal{M}_\Ptop} 
      \biggl(1+2\mathcal{M}_\Ptop \biggr)+ 
       \frac{\Kz^2 (1+a_\PW^2) (2 + \mathcal{M}_\Ptop)}
              {4\mathcal{S}_\PW^2(1-\mathcal{M}_\PZ)^2}
      + \frac{3 \Kz\Kg a_\PW e_\Pq}
                           {\mathcal{S}_\PW(1-\mathcal{M}_\PZ)} \biggr]
\end{equation} 
\noindent
in the limit of a massless c quark; $\Kg$ and $\Kz$ define the strength of the 
anomalous coupling for the current with a photon and a $\PZ$ boson,
respectively,
$s$ is the centre-of-mass energy squared, $\alpha$ is the fine structure 
constant, $e_{\rm{{q}}} =2/3$ and $m_\Ptop$ are the charge fraction and the 
mass of the top quark, $a_\PW = 1-4\sin^2 \theta_\PW$, 
$\mathcal{S}_\PW = \sin^22\theta_\PW$,
$\theta_\PW$ is the Weinberg angle, $\mathcal{M}_\Ptop = m^2_\Ptop/s$ and
$\mathcal{M}_\PZ = m^2_\PZ/s$. QCD corrections are taken into account
following the prescriptions of Reference~\citen{Reinders:1985}. Moreover,
the effect of initial state radiation (ISR) must be included.
Born-level and QCD+ISR corrected predictions for the cross section are shown
in Figure~\ref{fig:cross_sections}b.
The FCNC branching fraction limits set by CDF correspond to upper limits on
the anomalous couplings of
$\Kg^2 < 0.176$ and $\Kz^2 < 0.533$~\cite{Obraztsov:1998if}, 
at the 95\% confidence level.

\section*{Data and Monte Carlo Samples}
\label{sec:datamc}
\label{section:sample}
This study is based on 634~$\mbox{pb}^{-1}$ of data collected by the L3
detector~\cite{l3_1990_1} 
at LEP at $\sqrt{s} = 189 - 209$~GeV.
This integrated luminosity corresponds to the seven ranges of average 
centre-of-mass energies shown in Table~~\ref{tab:lumi}.
\renewcommand{\arraystretch}{1}
\begin{table}[htbp]
  \begin{center}
    \leavevmode
    \begin{tabular}{|ll|*{7}{c}|}
      \hline
        $<\sqrt{s}>$ &(GeV) 
        & 188.6 & 191.6 &  195.5 & 199.6 & 201.8 & 204.8 & 206.6 \\
      \hline
       Luminosity & (pb$^{-1}$) 
   & 176.8  
   &  \phantom{0}29.8  
   &  \phantom{0}84.1  
   &  \phantom{0}84.0  
   &  \phantom{0}37.7  
   &  \phantom{0}86.0  
   & 135.5  \\
      \hline
    \end{tabular}
    \icaption{\rm Average centre-of-mass energies and their corresponding 
      integrated luminosities.
     \label{tab:lumi}}
   \end{center}
 \end{table}
\par
Crucial to this analysis is the prediction of Standard Model backgrounds,
which rely on the following Monte Carlo (MC) programs:
PYTHIA~\cite{pythia_1} and KK2f~\cite{kk2f} for $\backqq$,
KORALW~\cite{koralw_1} for $\backww$, 
KORALZ~\cite{koralz} for $\backtt$, PHOJET~\cite{phojet_1}
for two-photon interactions and  
EXCALIBUR~\cite{excalibur} for other four-fermion final states.
Single top production MC events were generated with
a modified version of PYTHIA~\cite{jetset_modified}. 
QCD colour reconnection effects are taken into account in the framework of 
the Lund fragmentation model~\cite{Andersson:1983ia}, forcing the top decay 
$\Ptop\rightarrow\PWp\Pbottom$ before the fragmentation takes place.
\par
The response of the L3 detector is simulated using the 
GEANT~\cite{geant} program, taking into account the effects of 
multiple scattering, energy loss and showering in the detector. Hadronic
interactions in the detector are modeled using the 
GHEISHA~\cite{geant} program. Time dependent detector inefficiencies,
as monitored during the data taking period, are also simulated.

\section*{Analysis Procedures
\label{section:analysis}}
In the reaction $\eetc$ at \LEP centre-of-mass energies, 
the top quark is produced almost at rest and quickly decays via
$\Ptop\rightarrow\PWp\Pbottom$ without forming top-flavoured hadrons. 
Depending on the subsequent decay of the $\PW$ boson, the expected final 
state signatures are two jets, one lepton and missing energy
($\Ptop\Pacharm\rightarrow\PWp\Pbottom\Pacharm
\rightarrow\rm{l}^+\nu\Pbottom\Pacharm$) or
four jets 
($\Ptop\Pacharm\rightarrow\PWp\Pbottom\Pacharm
\rightarrow\Pq\Paq'\Pbottom\Pacharm$),
hereafter referred to as the leptonic and hadronic channels, respectively.
In both channels, the c-quark energy $E_\Pcharm$ has a fixed value for a 
given $\sqrt{s}$:
\begin{eqnarray}
E_\Pcharm & = & \frac{\sqrt{s}}{2}\biggl(1 - \frac{m_\Ptop^2}{s}\biggr),
\label{eq:ec}
\end{eqnarray}
whereas the b-quark energy $E_\Pbottom$ has an almost fixed value which does not
depend on $\sqrt{s}$, in the limit of a top quark at rest in the centre-of-mass
frame:
\begin{eqnarray}
E_\Pbottom & \simeq & \frac{m_\Ptop}{2}
\biggl(1 - \frac{m_\PW^2}{m_\Ptop^2}\biggr),
\label{eq:eb}
\end{eqnarray}
where $m_\PW$ is the mass of the $\PW$ boson.
After hadronisation, the c and b quarks yield jets with almost fixed energies.
Since both channels are characterised by the presence of a b quark, a clear 
signature exists and is exploited by using a b-jet tagging 
algorithm~\cite{l3_1997_18}, which is mainly based on lifetime information.
\par
Standard search procedures are applied to both the leptonic and hadronic 
channels. First, a preselection is applied which significantly reduces the
background while keeping a high signal efficiency; this is especially 
effective  against background from low multiplicity events and from 
two-photon interactions. Then, a further set of channel-specific selection 
criteria is chosen to increase the signal-to-background ratio. Finally,
a discriminating variable is built using a neural network technique.

\section*{Leptonic Channel
\label{subsection:leptonic}}
The signature for the leptonic channel is one energetic lepton,
large missing momentum and two jets with a
large difference in energy. The most energetic jet is assumed to stem from
the hadronisation of the b quark.
The $\backww$ and $\backqq$ processes constitute the main background.
\par
A preselection is applied requiring events to have at least three tracks, 
more than 15 calorimetric clusters and a visible energy greater than
$0.25\times \sqrt{s}$, but less than $0.9\times \sqrt{s}$.
The presence of a well identified lepton is required; if there is more than 
one reconstructed lepton, the most energetic one is retained.
An electron candidate is identified as a track with an
associated cluster in the electromagnetic calorimeter;
a muon candidate is reconstructed as a track in the central tracker matched 
to one in the muon spectrometer; a tau lepton is identified as a 
low-multiplicity jet. 
All clusters in the event, except the ones associated with the 
reconstructed lepton, are combined to form two hadronic jets
using the DURHAM algorithm~\cite{durham1}. The jet axes and the
missing momentum vector must be at least 15$^\circ$ and 26$^\circ$,
respectively, away from the beam axis.\par
\begin{figure}
\begin{eqnarray*}
  \epsfig{file=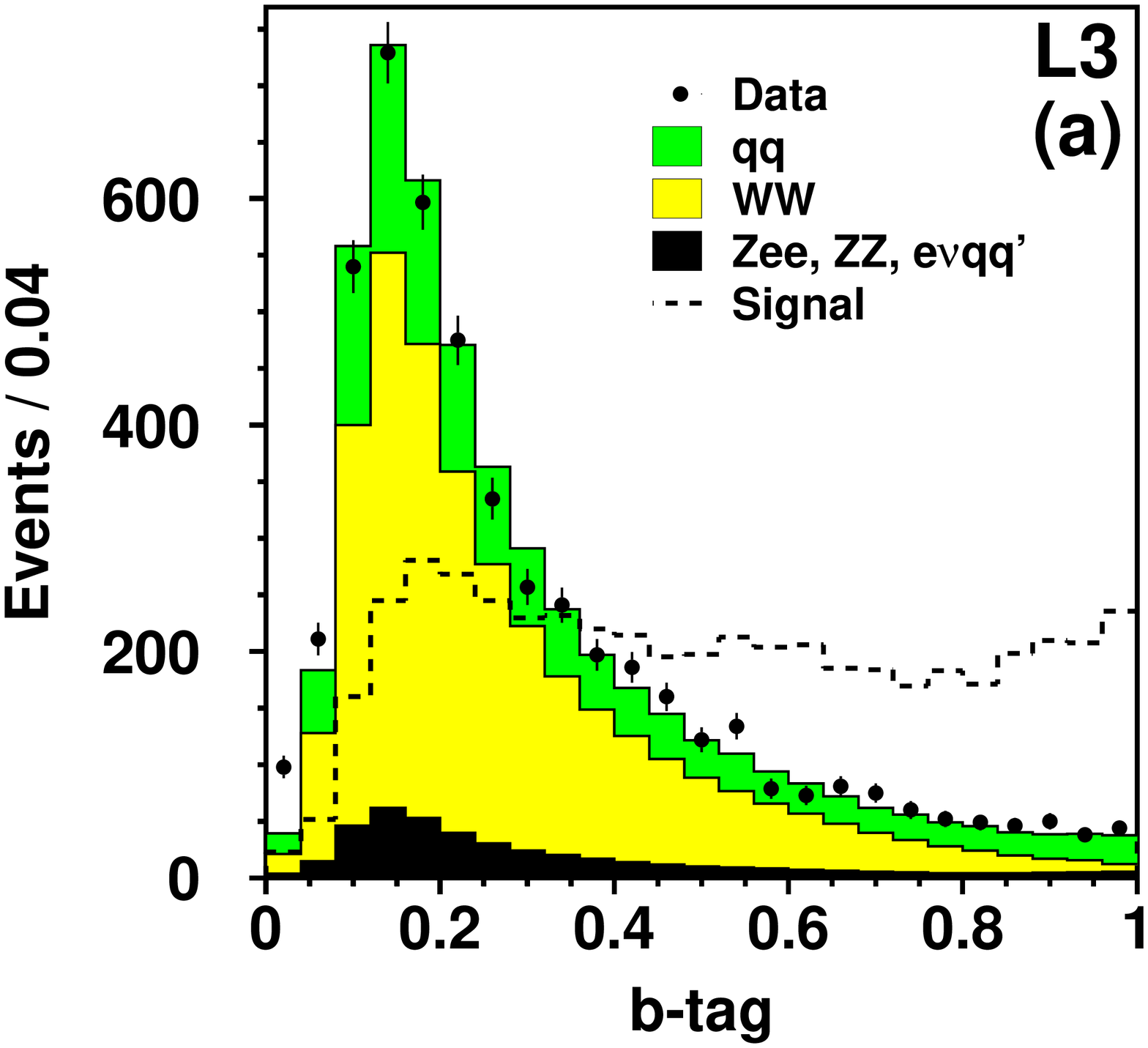,width=7.9cm} &
  \epsfig{file=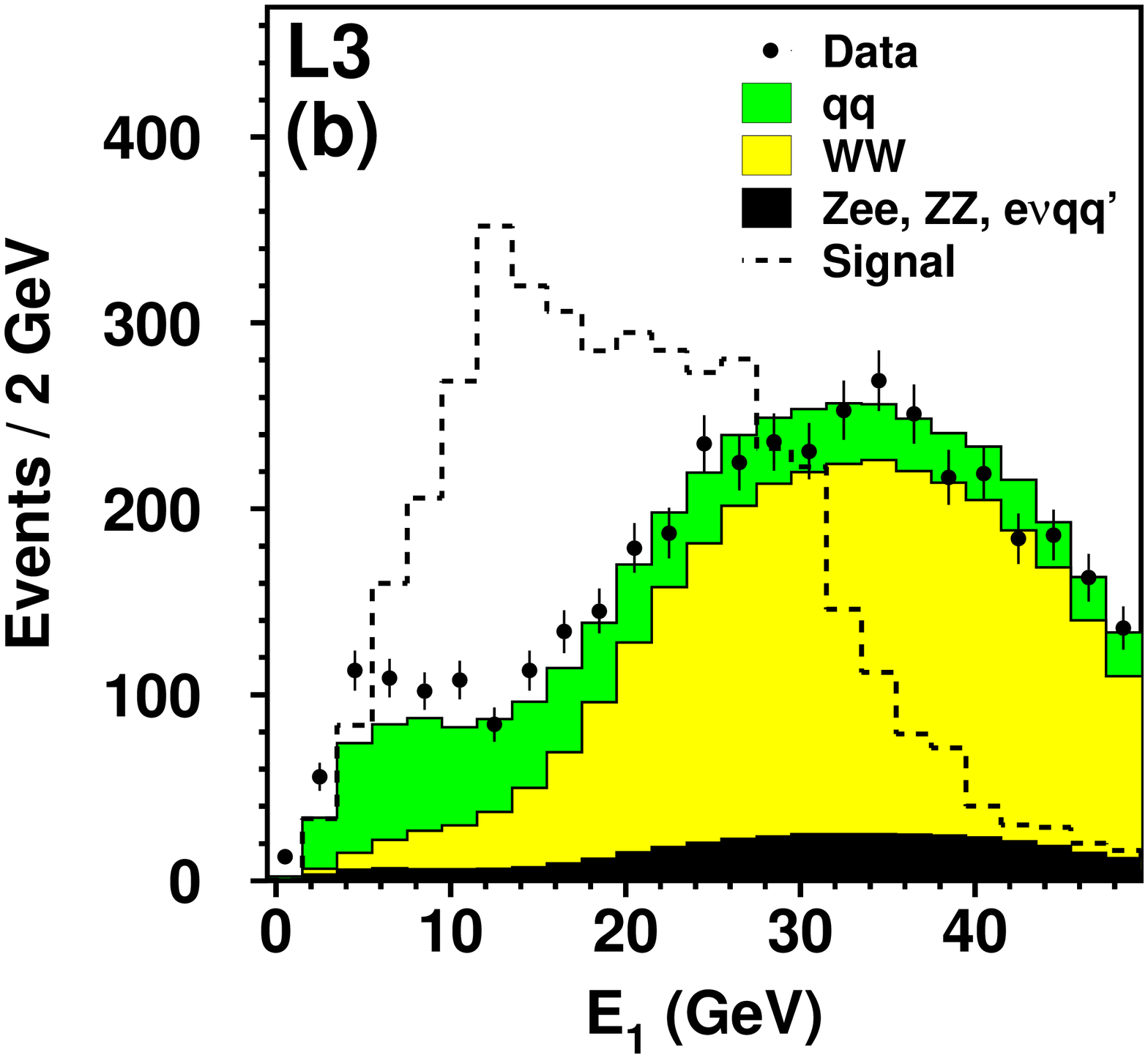,width=7.9cm}
\end{eqnarray*}
  \icaption{\rm Distributions for the leptonic channel of (a) the b-tag 
   discriminant variable for the most energetic jet, and (b) the energy 
   $E_1$ of the least energetic jet.
   The signal histogram is for a top quark mass of 175~GeV and is normalised 
   to the number of data events. Background expectations are also shown.
    \label{fig:leptonic_plots}
    }
\end{figure}

To further reject background events, the energy of the lepton is
required to be at least 10~GeV. The energy of the most energetic jet is 
required to exceed 60~GeV, whereas an upper bound is set on the energy $E_1$
of the least energetic
jet, as detailed in Table~\ref{tab:lcuts}. The width of the least energetic 
jet must be less than 0.4, the jet width being defined as the scalar sum of 
the transverse momenta of the jet clusters, normalised to the jet energy.
The missing momentum is required to exceed 25~GeV, the lepton plus missing 
momentum invariant mass must be larger than 20~GeV and the two-jet invariant 
mass $M_{\Pq\Paq}$ has to lie outside a window around the $\PW$
mass, as given in Table~\ref{tab:lcuts}.
The distributions of some relevant
variables are shown in Figure~\ref{fig:leptonic_plots}.
From the overall selection, 346 events are left in the
data sample, with $357.0\pm 1.8$ expected from Standard Model backgrounds.
The signal efficiency is 10.6\% as detailed in Table~\ref{tab:stat}
which also gives results at all centre-of-mass energies.
\renewcommand{\arraystretch}{1}
  \begin{table}[b]
    \begin{center}
      \leavevmode
      \begin{tabular}{*{8}{c}|}
        \hline
        \multicolumn{1}{|l|}{$\sqrt{s}$ (GeV)}
        & 188.6 & 191.6 &  195.5 & 199.6 & 201.8 & 204.8 & 206.6 \\
        \hline
        \multicolumn{1}{|l|}{$E_1$ (GeV) $\le$} &
        17.0 & 21.0 & 23.0 & 27.0 & 30.0 & 34.0 & 34.0\\
        \multicolumn{1}{|l|}{$|M_{\Pq\Paq} - m_\PW|$ (GeV) $\ge$} & 
        14.0 & 11.0 & \phantom{0}8.5 & \phantom{0}6.0 & \phantom{0}6.0 & 
        \phantom{0}3.0 & \phantom{0}3.0\\
        \hline
      \end{tabular}
      \icaption{\rm Centre-of-mass energy dependent cuts used for 
        the leptonic channel.
        \label{tab:lcuts}
        }
    \end{center}
  \end{table}
In order to enhance the separation between signal
and background, a neural network technique~\cite{jetnet_1}
is then employed. The most important event variables
are used as inputs to a neural
network with 10 input nodes and two output nodes,
$\mathcal{O}_{tc}$, $\mathcal{O}_{back}$,
corresponding to the signal and the background,
respectively. The input variables are related to the magnitude and 
direction of the 
missing momentum vector, the b-tag value, the invariant mass of the two-jets
system and the invariant mass of the 
lepton plus missing-momentum system.
The final discriminating variable is then
obtained as the product $ \mathcal{O}= \mathcal{O}_{tc} \times 
(1-\mathcal{O}_{back})$.
\par\smallskip
\renewcommand{\arraystretch}{1}
  \begin{table}[h]
    \begin{center}
      \leavevmode
      \begin{tabular}{|*{7}{c|}}
\hhline{~*{6}{-}}
\multicolumn{1}{c|}{}& \multicolumn{3}{|c|}{Leptonic channel} & 
\multicolumn{3}{|c|}{Hadronic channel}\\
\hhline{*{1}{-}*{6}{-}}
$\sqrt{s}$~(GeV) & Data & Back. & Sig. eff. (\%) & Data & Back. & Sig. eff. (\%) \\
\hline
 188.6 &  
   \phantom{0}10 & \phantom{0}13.6 $\pm$   0.3 & 10.7 $\pm$  0.2 &
   \phantom{0}95 &  \phantom{0}76.6 $\pm$  0.8 & 21.0 $\pm$  0.4 \\

 191.6 &  
    \phantom{10}3 & \phantom{10}4.6 $\pm$   0.2 & 11.1 $\pm$  0.3 &
   \phantom{0}14 &  \phantom{0}14.3 $\pm$  0.5 & 23.1 $\pm$  0.6 \\

 195.5 &  
   \phantom{0}23 & \phantom{0}21.1 $\pm$   0.4 & 10.4 $\pm$  0.3 &
   \phantom{0}43 &  \phantom{0}38.7 $\pm$  1.0 & 22.1 $\pm$  0.6 \\

 199.6 &  
   \phantom{0}35 & \phantom{0}40.5 $\pm$   0.4 & 10.7 $\pm$  0.3 &
   \phantom{0}37 &  \phantom{0}38.1 $\pm$  0.5 & 21.5 $\pm$  0.6 \\

 201.8 &  
   \phantom{0}22 & \phantom{0}23.7 $\pm$   0.2 & 10.2 $\pm$  0.3 &
   \phantom{0}19 &  \phantom{0}19.4 $\pm$  0.4 & 21.6 $\pm$  0.6 \\

 204.8 &  
  104 &  \phantom{0}99.1 $\pm$   1.1 & 10.5 $\pm$  0.3 &
   \phantom{0}50 &  \phantom{0}41.1 $\pm$  0.7 & 20.7 $\pm$  0.6 \\

 206.6 &  
  149 & 154.4 $\pm$   0.9 & 10.5 $\pm$  0.3 &
   \phantom{0}63 &  \phantom{0}59.7 $\pm$  0.6 & 20.2 $\pm$  0.5 \\
\hline
   All &   
   346  & 357.0 $\pm$ 1.8 & 10.6 $\pm$ 0.1 &
   321 & 287.9 $\pm$ 1.6 & 21.1 $\pm$ 0.2 \\
\hline
      \end{tabular}
      \icaption{\rm The number of data events for each centre-of-mass
   energy range, the expected number of background events and the
   Monte Carlo predicted signal efficiencies, for both the leptonic
   and hadronic channels. The errors are statistical only. 
   The signal efficiency is for a top quark mass of 175~GeV. 
   The corresponding W decay branching fraction is taken into account 
   in the signal efficiency.
        \label{tab:stat}
        }
    \end{center}
  \end{table}

\section*{Hadronic Channel
\label{subsection:hadronic}}
The signature for the hadronic channel is four hadronic jets, two jets having
an almost fixed energy and two being the decay products of a W. In
addition, one jet must have a strong b-jet signature.
The $\backww$ and $\backqq$ processes constitute the main background.
\par
A hadronic preselection, which requires
a visible energy of at least $0.7 \times \sqrt{s}$ and an effective 
centre-of-mass
energy larger than $0.85 \times \sqrt{s}$, is applied. 
The effective centre-of-mass energy is computed after having removed photon
radiation in the initial state.
We also require 
more than 20 reconstructed tracks and more than two jets, built using the 
DURHAM algorithm with a resolution parameter $y_{cut}$~=~0.02. 
\par
The events are forced into a four-jet topology by changing the value of 
$y_{cut}$.
The c-jet candidate is defined as the jet whose energy
is closest to the value expected from Equation~(\ref{eq:ec}) for 
$m_{\rm t} = 174.3$~GeV.
Using the remaining three jets, the $\PW$ boson is identified as
the jet pair with invariant mass closest to the nominal $\PW$ mass. 
The remaining jet is assumed to be the b jet.\par
By comparing the direction of the leading c quark in MC events with the c-jet 
defined above, it is observed that the c selection purities are about
49\%, 53\% and 60\%  for $\sqrt{s} =$~186.6, 191.6 and 195.5 GeV, respectively,
and approach the limit of 62\% for $\sqrt{s} \ge$~201.8~GeV.
\par
\begin{figure}
\begin{eqnarray*}
  \epsfig{file=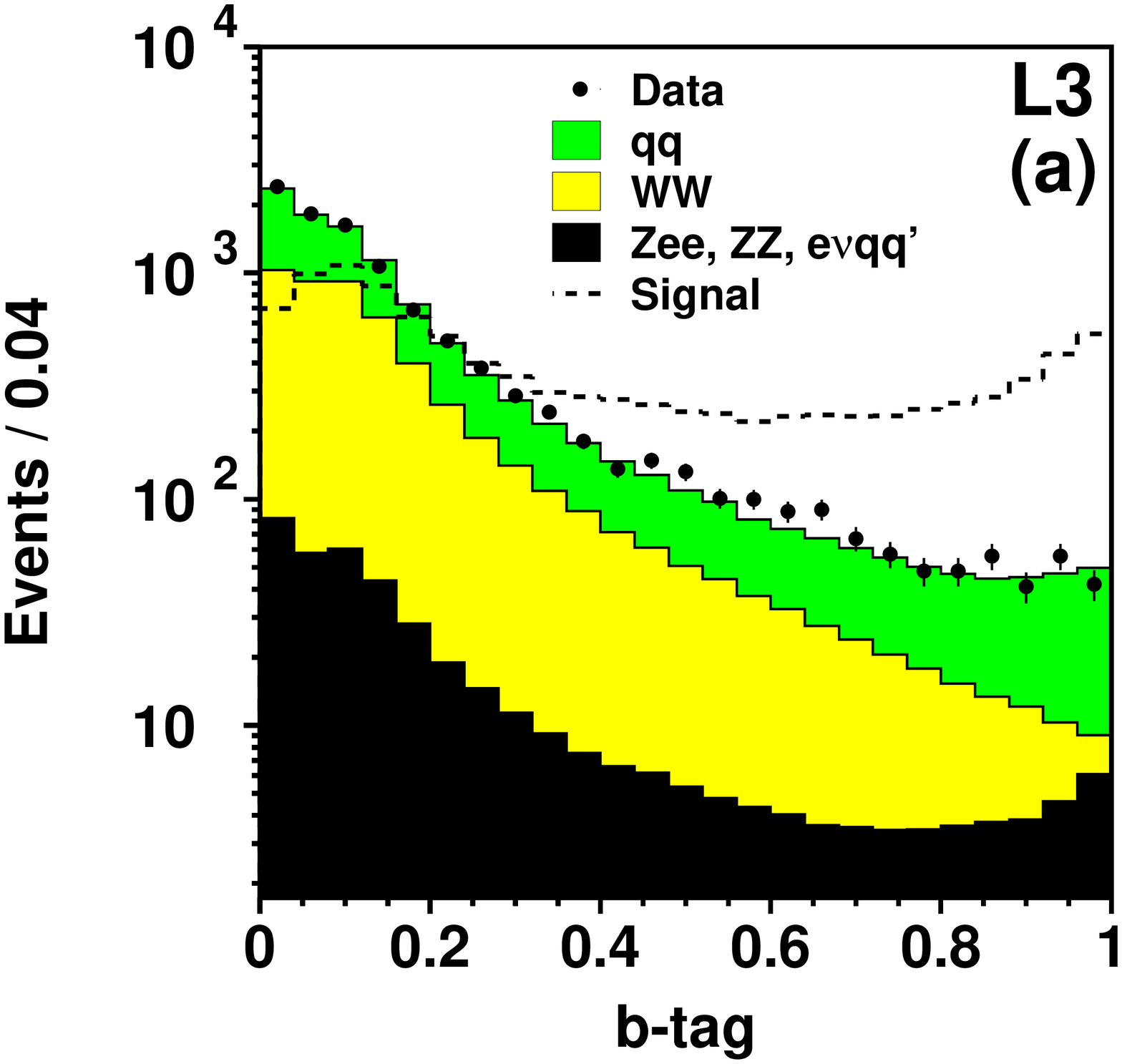,width=7.9cm} &
  \epsfig{file=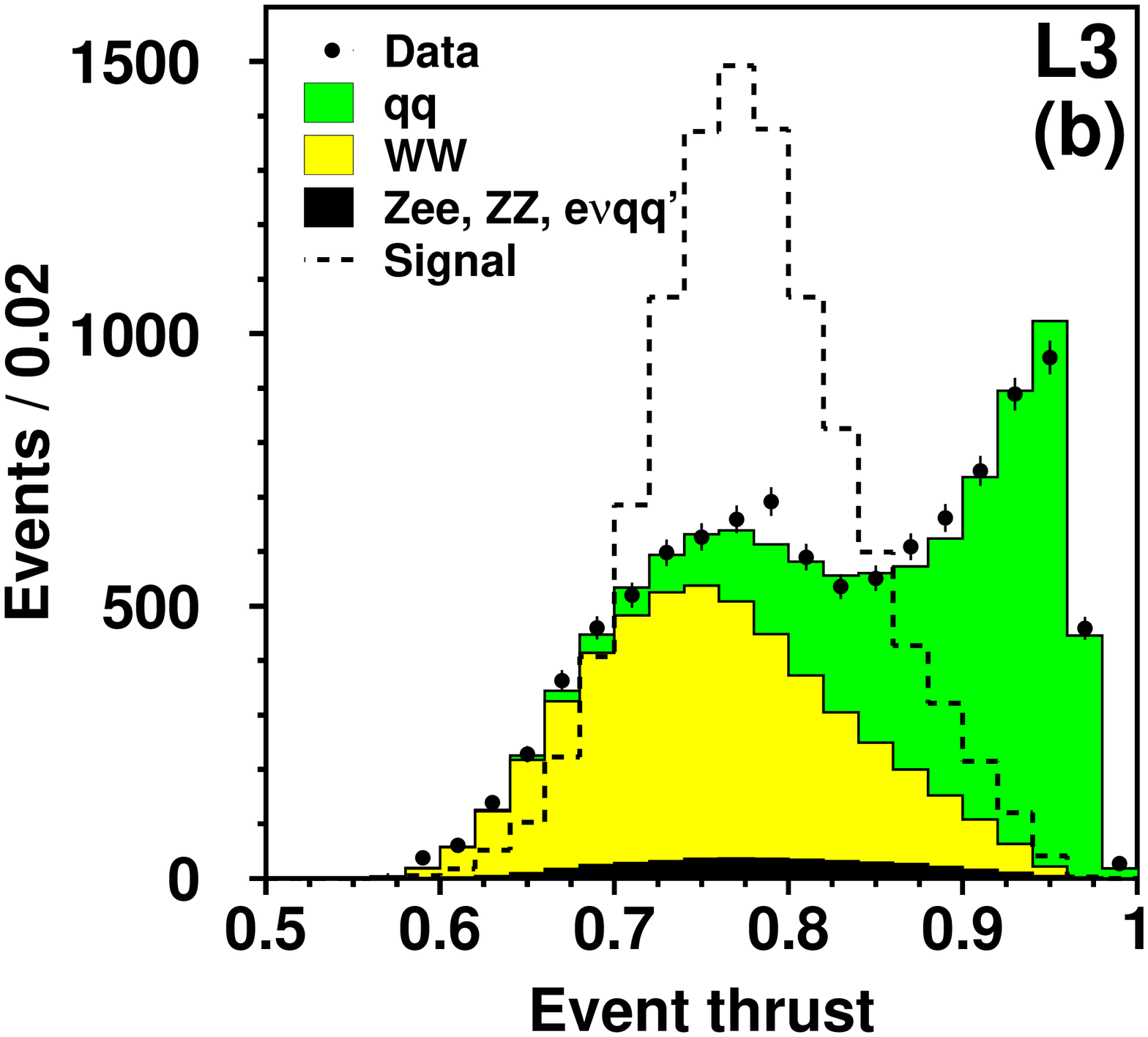,width=7.9cm} \\
  \epsfig{file=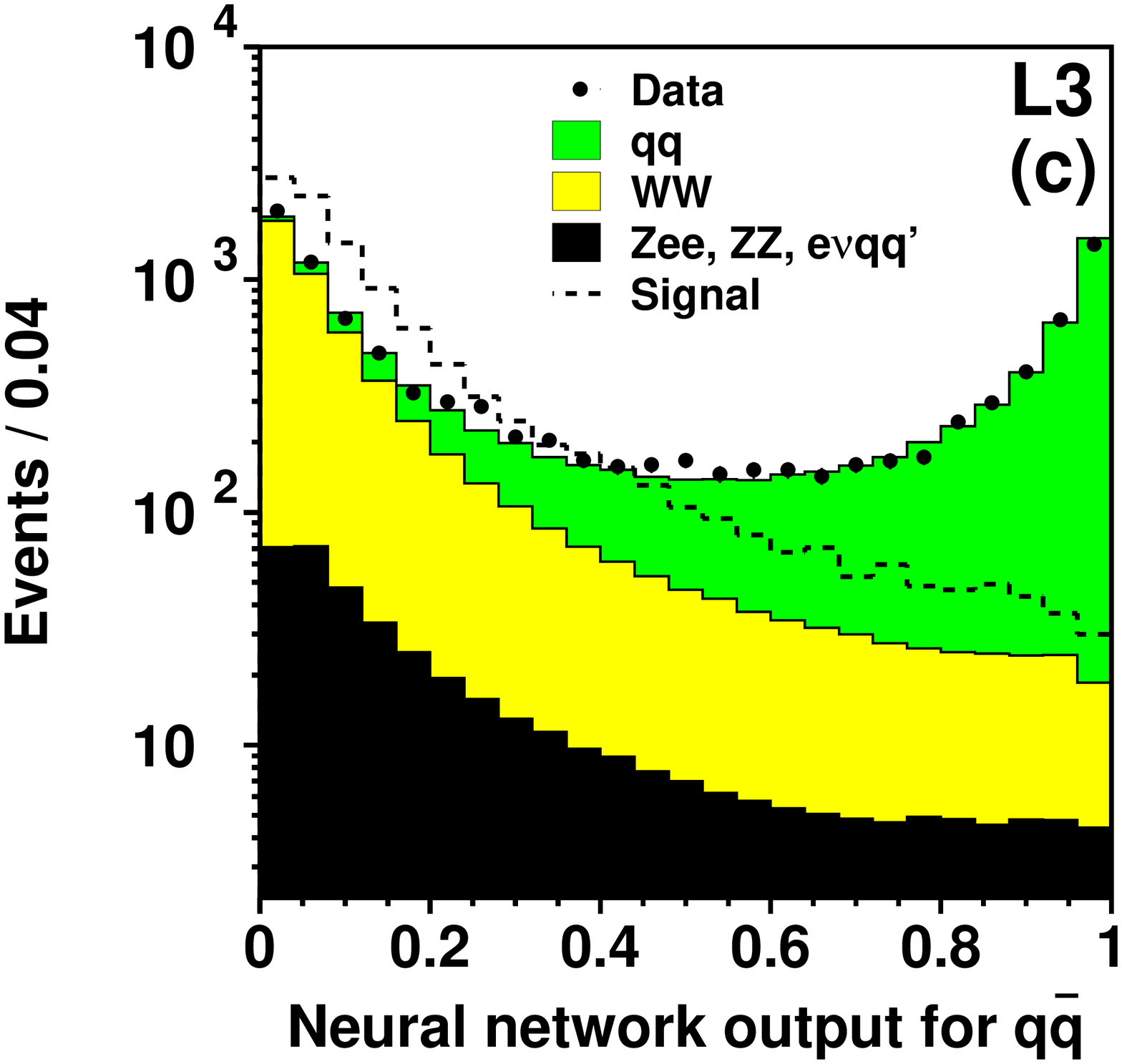,width=7.9cm} &
  \epsfig{file=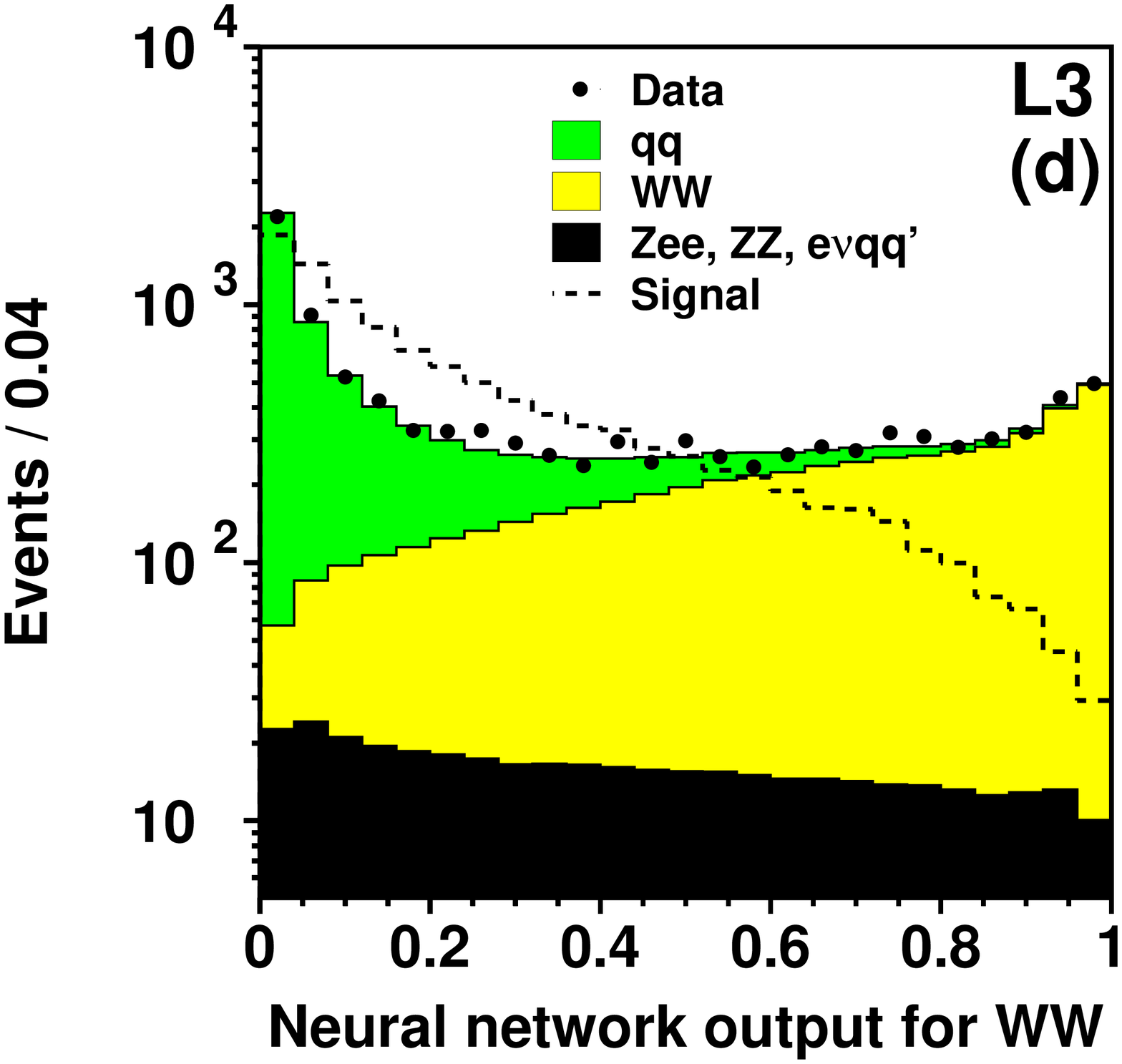,width=7.9cm}
\end{eqnarray*}
  \icaption{\rm
   Distributions for the hadronic channel after preselection of (a) the b-tag
   variable of the b jet, (b) the event thrust and (c), (d) the neural network
   outputs related to $\backqq$ and $\backww$. 
   The signal histogram is for a top quark mass of 175~GeV and is normalised 
   to the number of data events. Background expectations are also shown.
    \label{fig:hadronic_plots}
    }
\end{figure}

A neural network with 24 input nodes and three output nodes is then used.
The neural network input variables are related to such jet 
characteristics as their energies and masses, the b-tag discriminant 
of the b jet candidate
and various event shape variables. Some of these distributions are shown in
Figures~\ref{fig:hadronic_plots}a and \ref{fig:hadronic_plots}b.
One network output selects the signal and is used as the final discriminating 
variable. The other two outputs, which tag $\backqq$ and $\backww$ events,
are used 
to further reject the background by applying a cut of 0.1 and 0.3,
respectively. Their distributions are shown in 
Figures~\ref{fig:hadronic_plots}c and \ref{fig:hadronic_plots}d. 
After all cuts, the final sample consists of 321 data events, to be compared
with $287.9 \pm 1.6$ expected background events, and a
21.1\% signal efficiency, as detailed in Table~\ref{tab:stat}.

\section*{Study of Systematic Uncertainties}
\label{section:systematics}
The search for single top production discussed in this Letter relies
heavily on the comparison between the data and its associated Monte
Carlo simulations.  Uncertainties in these simulations give rise to
three sources of systematic uncertainties whose effects on the 
single top cross section are shown in Table 4.  First, the finite
statistics of the Monte Carlo used for the signal and background
simulations affect the determination of the signal efficiency and the
level of background contamination.  Secondly, the background 
cross sections are fixed in the interpretation
of the observed events in terms of the single top
cross section.
The effects of a variation of the $\backqq$, $\backww$ and $\backzz$ cross
sections of 2\%, 0.5\% and 10\%, respectively, are reported in 
Table~\ref{tab:syst}.
Thirdly, there can be differences between the actual and simulated
detector performance, affecting the description of variables used as
inputs to the neural network.  In particular, we study the
variables used for the lepton identification, the global event shape,
the energy measurement and the b-tag.  Their effects on the single top cross 
section are also given in Table~\ref{tab:syst}, that also lists effects from 
the modeling of other variables.  Finally, the systematic due to uncertainties 
in the measurement of the integrated luminosity is negligible.
\renewcommand{\arraystretch}{1}
\begin{table}[h]
\begin{center}
\leavevmode
\begin{tabular}{|l|rl|}
\hline
\multicolumn{1}{|c|}{Source} & 
\multicolumn{2}{c|}{$\Delta\sigma/\sigma$ (\%)} \\
\hline
MC background statistics    &     & $0.4$ \\ 
MC signal statistics        &     & $1.7$ \\ 
$\backqq$ modeling         &     & $0.2$ \\ 
$\backww$ modeling         & $<$ & $0.1$ \\ 
$\backzz$ modeling         &     & $0.4$ \\ 
Lepton identification       &     & $0.1$ \\
Event shape                 &     & $0.9$ \\
Energy resolution           &     & $1.6$ \\
b-tagging                   &     & $1.4$ \\
Other variables             &     & $0.5$ \\
Luminosity                  & $<$ & $0.1$ \\ 
Signal angular distribution &     & $2.0$ \\ 
\hline
\end{tabular}
\caption{\rm Relative systematic uncertainties affecting the single top 
cross section.}
\label{tab:syst}
\end{center}
\end{table}
\par
In addition to these effects, uncertainties in the modeling of the
signal process can affect our results.  To quantify this, various
signal samples with different final-state angular distributions are
simulated, inside the limits allowed by the anomalous coupling
scenario, with the effects reported in Table 4.  Systematic uncertainties on the signal Monte Carlo statistic are
propagated to the signal efficiency. Uncertainties on the background
Monte Carlo statistic and modeling are accounted for by lowering the
background expectations accordingly. Finally, the uncertainty in the
detector description is propagated by repeating the analysis with the
distributions of the variables entering the neural networks smeared to
agree
with the distributions observed in data.
An additional source
of uncertainty could be the value of $m_{\rm t}$ used in the
simulation.  The low momenta available for the top system at our
centre-of-mass energies would imply a change in the event kinematics
and hence in the selection efficiency.  Rather than assigning a
systematic uncertainty from this source, all results are parametrised
in terms of $m_{\rm t}$.

\section*{Results}
\begin{figure}
\begin{eqnarray*}
  \epsfig{file=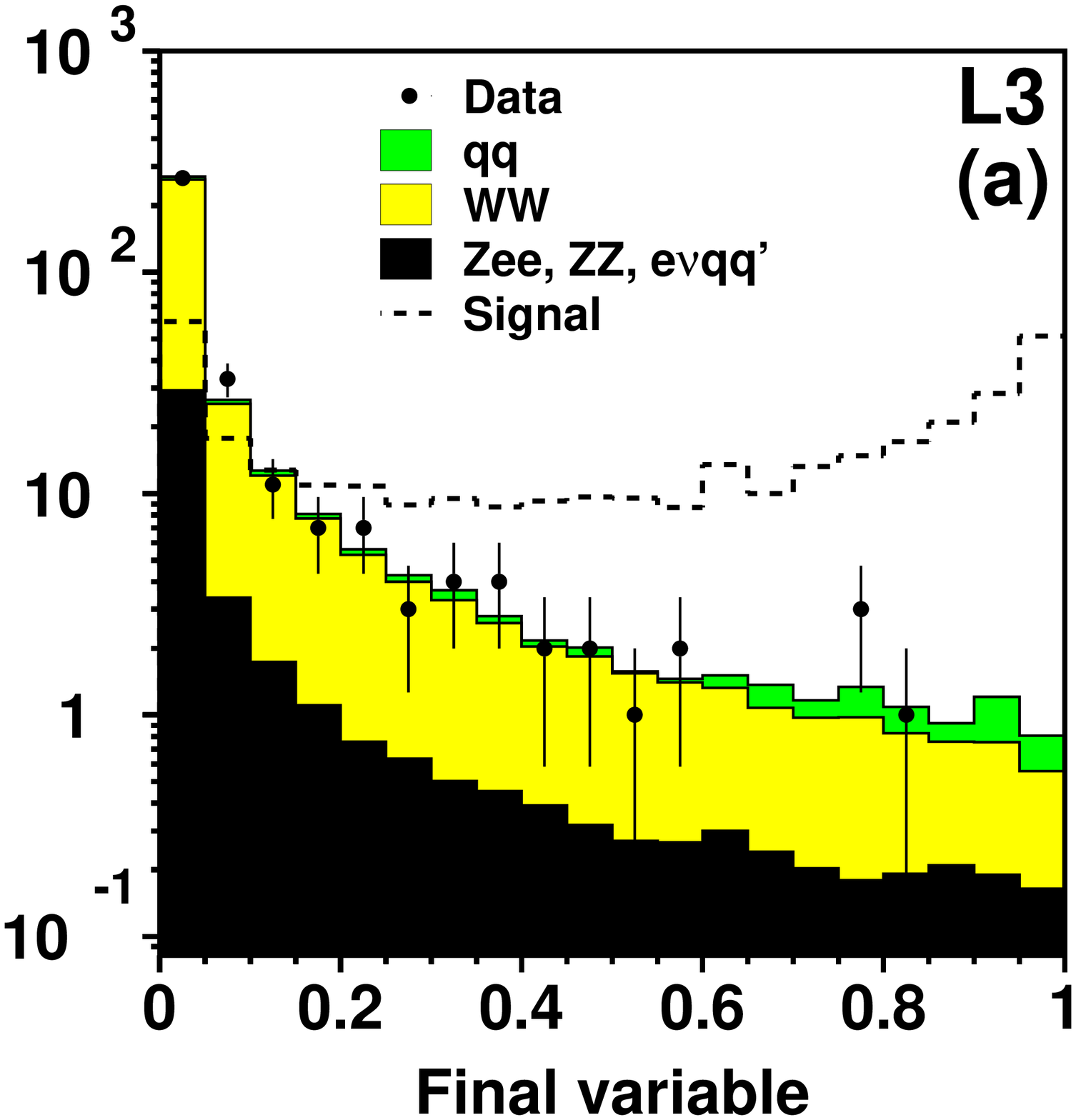,width=7.9cm} &
  \epsfig{file=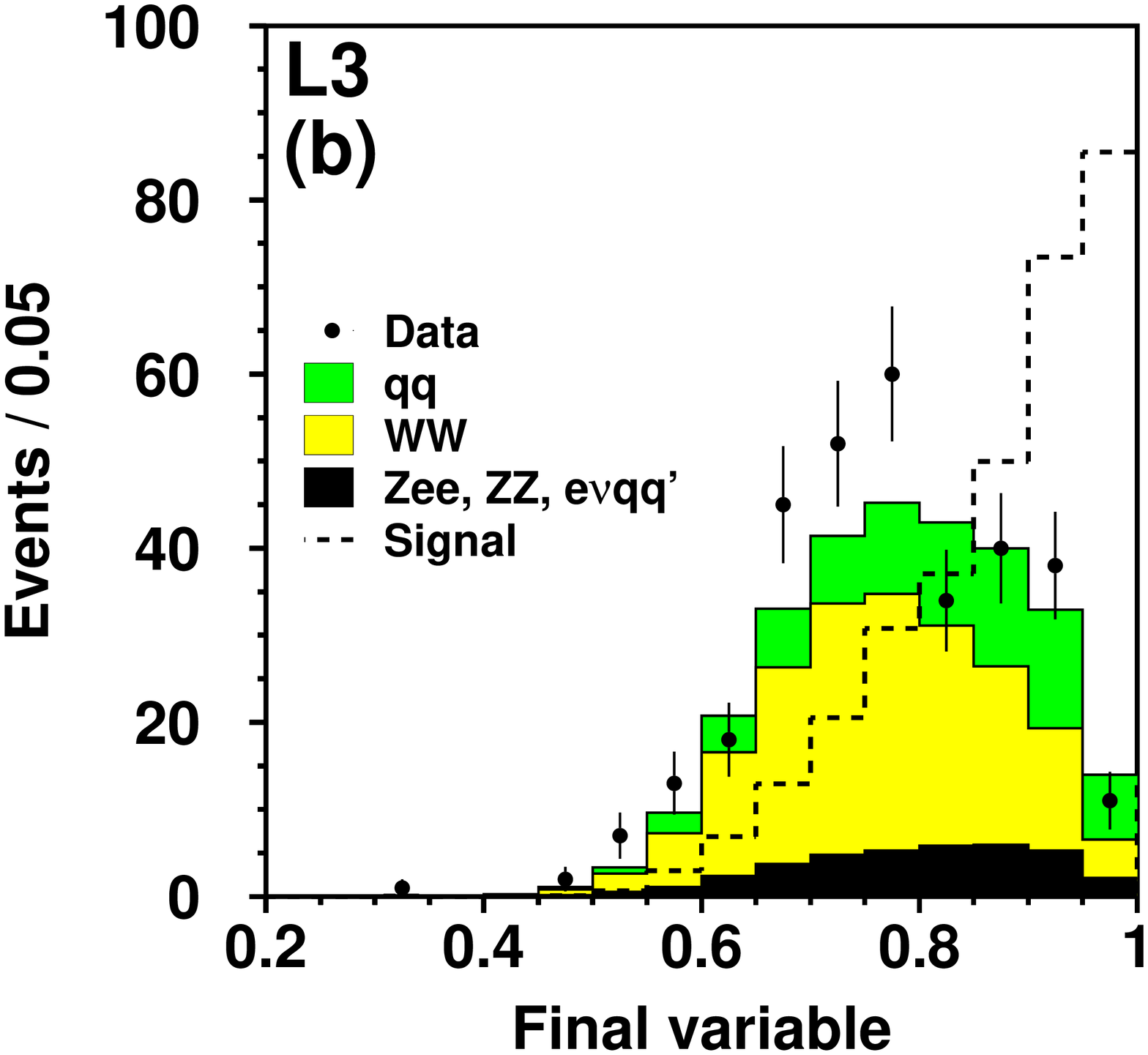,width=7.9cm} \\
\end{eqnarray*}
  \icaption{\rm
    Distribution of the final discriminating variable for (a) the leptonic
   channel and (b) the hadronic channel.
   The signal histogram is for a top quark mass of 175~GeV and is normalised 
   to the number of data events. Background expectations are also shown.
  \label{fig:final_plots}
    }
\end{figure}

Leptonic and hadronic final discriminating 
variables are shown in Figure~\ref{fig:final_plots}.
A very good discrimination between signal and background is 
achieved. No significant deviation from the Standard Model background 
expectation is observed.
Combined 95\% confidence level upper 
limits on the single top total cross section are 
derived~\cite{obraztsov,*new_method} and listed in 
Table~\ref{tab:xseclimits}.
For these limits, the branching ratio for the top decay is assumed to be
saturated by $\Ptop\rightarrow\PW\Pbottom$.
The limits are obtained for the signal process
$\eetc$, a deterioration of the limits of about 10\% is found for the 
corresponding process $\eetu$.\par
\begin{figure}
\begin{eqnarray*}
  \epsfig{file=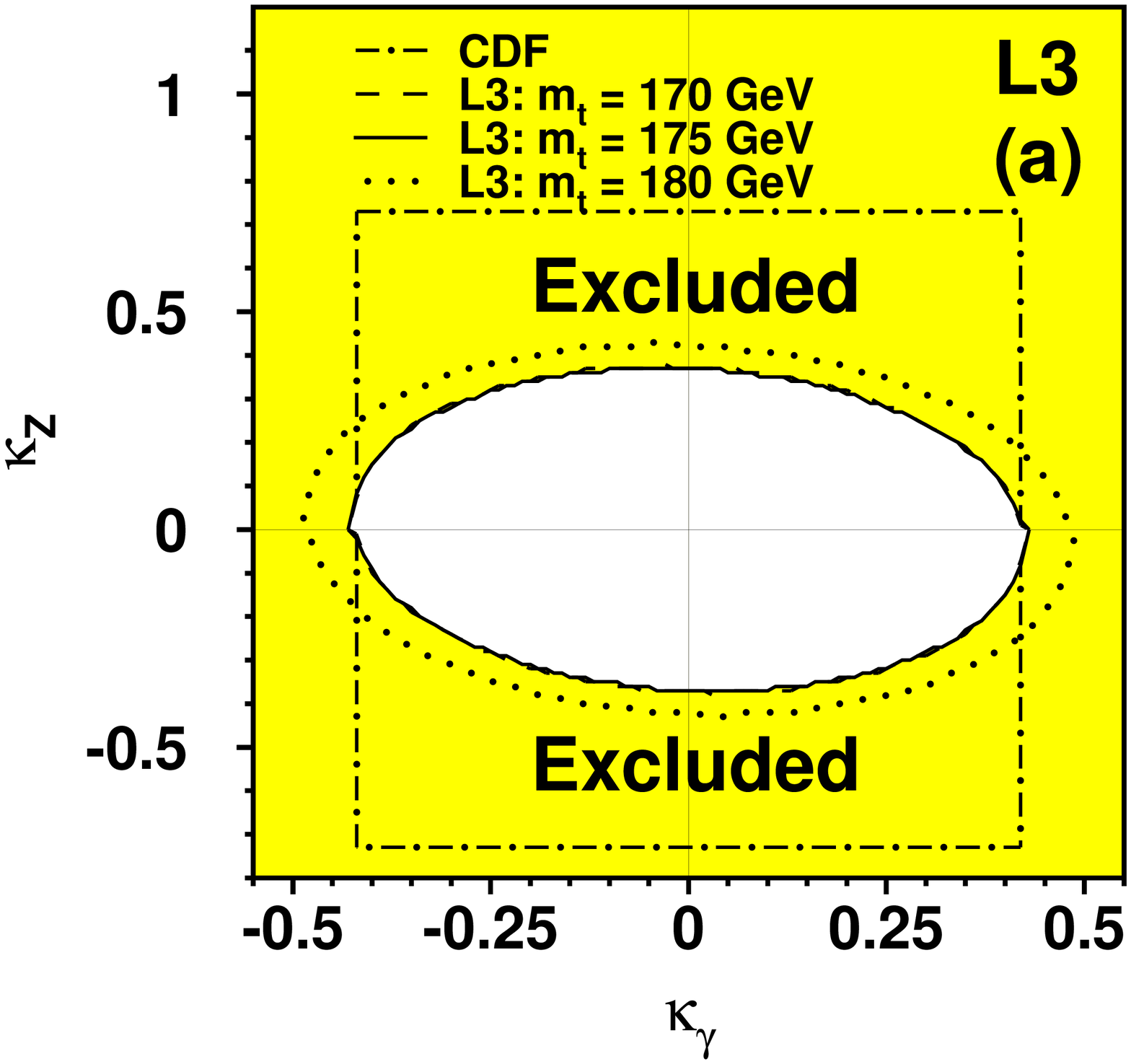,width=7.9cm} &
  \epsfig{file=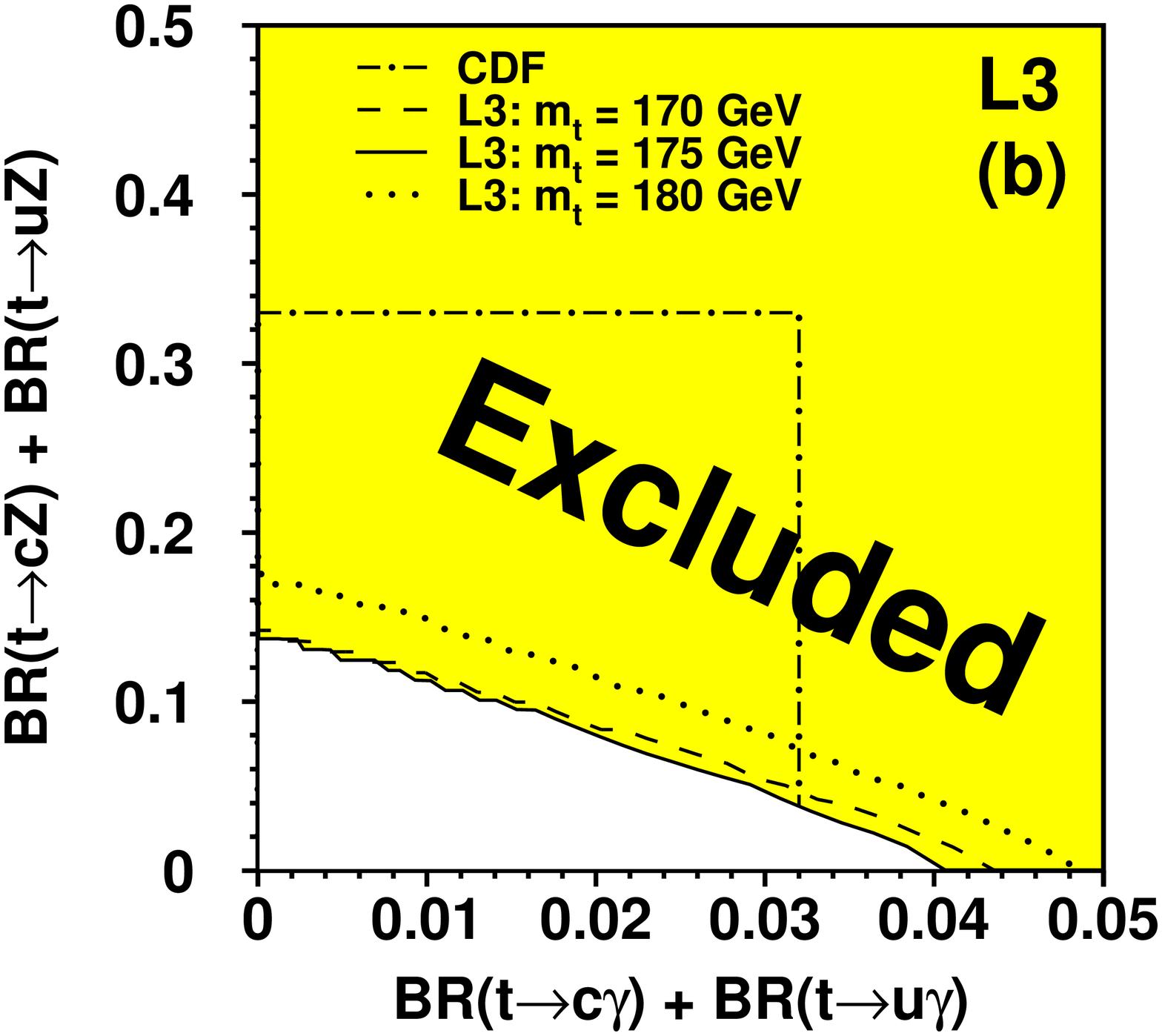,width=7.9cm}
\end{eqnarray*}
  \icaption{\rm
    Exclusion regions at the 95\% confidence level in
    (a) the $\Kz$ \textit{vs.} $\Kg$ plane and (b) the
    $\BR(\rm t \rightarrow \PZ \rm q)$ \textit{vs.} 
   $\BR(\rm t \rightarrow \gamma \rm q)$ plane
    for three different values of the
    top quark mass. The CDF exclusion domain is also shown.
    On Figure (a) the curves $m_{\rm t}$~=~170~GeV and 
    $m_{\rm t}$~=~175~GeV are almost overlapping.
  \label{fig:exclusion_plots}
    }
\end{figure}

Table~\ref{tab:lambda} lists the 95\% confidence level lower limit on the 
energy scale parameter $\Lambda$ of a possible [$\Ptop\Pacharm\Pep\Pem$] 
contact 
interaction, described in Equation~(\ref{eq:four}). 
Referring to the anomalous coupling formalism described in 
Reference~\citen{Obraztsov:1998if},
using the cross section expression of Equation~(\ref{eq:cross_section}), 
an exclusion region in the $\Kz$ \textit{vs.} $\Kg$ 
plane is obtained, as displayed in Figure~\ref{fig:exclusion_plots}a.
QCD and ISR corrections as well as 
flavour changing decays of the top quark through anomalous
vertices are taken into account in the limits computation.
Using the Born-level cross section of Equation~(\ref{eq:cross_section}),
a corresponding exclusion region in the 
$\BR(\Ptop\rightarrow\PZ\Pq)$ \textit{vs.} $\BR(\Ptop\rightarrow\gamma\Pq)$ 
plane is found, as shown in Figure~\ref{fig:exclusion_plots}b. 
The anomalous coupling formalism upper limits are summarised in 
Table~\ref{tab:kpar}.\par
In conclusion, no evidence for single top production at LEP is observed and
possible new physics responsible for this process is constrained.
\par\smallskip
\begin{table}[h]
  \begin{center}
    \leavevmode
    \begin{tabular}{|ll|*{7}{c|}}
      \hline
        $<\sqrt{s}>$ &(GeV) 
        & 188.6 & 191.6 &  195.5 & 199.6 & 201.8 & 204.8 & 206.6 \\
      \hline
       $\sigma_{95} (m_\Ptop = 170\GeV) $ & (pb) & 
  0.36 & 0.87 & 0.77 & 0.54 & 0.62 & 0.63 & 0.51 \\
       $\sigma_{95} (m_\Ptop = 175\GeV) $ & (pb) & 
  0.22 & 0.73 & 0.67 & 0.45 & 0.56 & 0.51 & 0.39 \\
       $\sigma_{95} (m_\Ptop = 180\GeV) $ & (pb) & 
  0.21 & 0.75 & 0.64 & 0.42 & 0.52 & 0.48 & 0.37 \\
      \hline
    \end{tabular}
    \icaption{\rm Measured 95\% confidence level upper limits, $\sigma_{95}$,
      on the total 
   cross section for single top production as a function of the
   centre-of-mass energy. 
   The limits are given for three assumptions on the top quark mass.
      \label{tab:xseclimits}
      }
   \end{center}
 \end{table}
\renewcommand{\arraystretch}{1}
\begin{table}[h]
\begin{center}
\begin{tabular}{|*{4}{c|}}
  \hhline{~*{3}{-}}
  \multicolumn{1}{c|}{}
   & \multicolumn{3}{|c|}{$\Lambda$ (TeV)}\\
  \hhline{~*{3}{-}}
  \multicolumn{1}{c|}{}
   & Vector coupling
   & Scalar coupling
   & Tensor coupling \\
  \hline
  $m_{\rm t}$ = 170 GeV & 0.76 & 0.65 & 1.24 \\
  $m_{\rm t}$ = 175 GeV & 0.75 & 0.65 & 1.24 \\
  $m_{\rm t}$ = 180 GeV & 0.70 & 0.60 & 1.16 \\
  \hline
\end{tabular}
\icaption{\rm 
Measured 95\% confidence level lower limits on the energy scale
parameter $\Lambda$  in TeV. Three different scenarios for the coupling
constants are considered: vectorial ($V_{ij}=1$, $S_{RR}=0$, $T_{RR}=0$),
scalar ($V_{ij}=0$, $S_{RR}=1$, $T_{RR}=0$) and tensorial
($V_{ij}=0$, $S_{RR}=0$, $T_{RR}=1$).
Limits are given for three values of the top quark mass.
\label{tab:lambda}
}
\end{center}
\end{table}
\begin{table}[h]
  \begin{center}
    \leavevmode
    \begin{tabular}{|c|r|r|r|}
      \hline
      \multicolumn{1}{|c|}{$m_{\rm t}$ (GeV)} & 
      \multicolumn{1}{|c|}{170} & \multicolumn{1}{|c|}{175} & 
    \multicolumn{1}{|c|}{180} \\
      \hline
      $|\Kz|$ & 0.38 & 0.37 & 0.43 \\
      $|\Kg|$ & 0.43 & 0.43 & 0.49 \\
      \hline
      $\BR(\Ptop\rightarrow\PZ\Pq)$ & 
        13.6\%  &  13.7\%  &  17.0\% \\
      $\BR(\Ptop\rightarrow\gamma\Pq)$ & 
       \phantom{0}4.4\% & \phantom{0}4.1\% & \phantom{0}4.9\% \\
      \hline
    \end{tabular}
    \icaption{\rm Measured 95\% confidence level upper limits on
   the anomalous-coupling parameters $\Kz$ and $\Kg$ and on the FCNC top 
   decay branching fractions. 
   Limits are given for three values of the top quark mass.
      \label{tab:kpar}
      }
  \end{center}
\end{table}

\newpage
\typeout{   }     
\typeout{Using author list for paper 256 -  }
\typeout{$Modified: Jul 15 2001 by smele $}
\typeout{!!!!  This should only be used with document option a4p!!!!}
\typeout{   }
%
%
%
%
%
%

\newcount\tutecount  \tutecount=0
\def\tutenum#1{\global\advance\tutecount by 1 \xdef#1{\the\tutecount}}
\def\tute#1{$^{#1}$}
\tutenum\aachen            
\tutenum\nikhef            
\tutenum\mich              
\tutenum\lapp              
\tutenum\basel             
\tutenum\lsu               
\tutenum\beijing           
\tutenum\berlin            
\tutenum\bologna           
\tutenum\tata              
\tutenum\ne                
\tutenum\bucharest         
\tutenum\budapest          
\tutenum\mit               
\tutenum\panjab            
\tutenum\debrecen          
\tutenum\dublin            
\tutenum\florence          
\tutenum\cern              
\tutenum\wl                
\tutenum\geneva            
\tutenum\hefei             
\tutenum\lausanne          
\tutenum\lyon              
\tutenum\madrid            
\tutenum\florida           
\tutenum\milan             
\tutenum\moscow            
\tutenum\naples            
\tutenum\cyprus            
\tutenum\nymegen           
\tutenum\caltech           
\tutenum\perugia           
\tutenum\peters            
\tutenum\cmu               
\tutenum\potenza           
\tutenum\prince            
\tutenum\riverside         
\tutenum\rome              
\tutenum\salerno           
\tutenum\ucsd              
\tutenum\sofia             
\tutenum\korea             
\tutenum\purdue            
\tutenum\psinst            
\tutenum\zeuthen           
\tutenum\eth               
\tutenum\hamburg           
\tutenum\taiwan            
\tutenum\tsinghua          

{
\parskip=0pt
\noindent
{\bf The L3 Collaboration:}
\ifx\selectfont\undefined
 \baselineskip=10.8pt
 \baselineskip\baselinestretch\baselineskip
 \normalbaselineskip\baselineskip
 \ixpt
\else
 \fontsize{9}{10.8pt}\selectfont
\fi
\medskip
\tolerance=10000
\hbadness=5000
\raggedright
\hsize=162truemm\hoffset=0mm
\def\r{\rlap,}
\noindent

P.Achard\r\tute\geneva\ 
O.Adriani\r\tute{\florence}\ 
M.Aguilar-Benitez\r\tute\madrid\ 
J.Alcaraz\r\tute{\madrid,\cern}\ 
G.Alemanni\r\tute\lausanne\
J.Allaby\r\tute\cern\
A.Aloisio\r\tute\naples\ 
M.G.Alviggi\r\tute\naples\
H.Anderhub\r\tute\eth\ 
V.P.Andreev\r\tute{\lsu,\peters}\
% ...
\newpage
\begin{list}{A}{\itemsep=0pt plus 0pt minus 0pt\parsep=0pt plus 0pt minus 0pt
                \topsep=0pt plus 0pt minus 0pt}
\item[\aachen]
 I. Physikalisches Institut, RWTH, D-52056 Aachen, Germany$^{\S}$\\
 III. Physikalisches Institut, RWTH, D-52056 Aachen, Germany$^{\S}$
\item[\nikhef] National Institute for High Energy Physics, NIKHEF, 
     and University of Amsterdam, NL-1009 DB Amsterdam, The Netherlands
\item[\mich] University of Michigan, Ann Arbor, MI 48109, USA
\item[\lapp] Laboratoire d'Annecy-le-Vieux de Physique des Particules, 
     LAPP,IN2P3-CNRS, BP 110, F-74941 Annecy-le-Vieux CEDEX, France
\item[\basel] Institute of Physics, University of Basel, CH-4056 Basel,
     Switzerland
\item[\lsu] Louisiana State University, Baton Rouge, LA 70803, USA
\item[\beijing] Institute of High Energy Physics, IHEP, 
  100039 Beijing, China$^{\triangle}$ 
\item[\berlin] Humboldt University, D-10099 Berlin, Germany$^{\S}$
\item[\bologna] University of Bologna and INFN-Sezione di Bologna, 
     I-40126 Bologna, Italy
\item[\tata] Tata Institute of Fundamental Research, Mumbai (Bombay) 400 005, India
\item[\ne] Northeastern University, Boston, MA 02115, USA
\item[\bucharest] Institute of Atomic Physics and University of Bucharest,
     R-76900 Bucharest, Romania
\item[\budapest] Central Research Institute for Physics of the 
     Hungarian Academy of Sciences, H-1525 Budapest 114, Hungary$^{\ddag}$
\item[\mit] Massachusetts Institute of Technology, Cambridge, MA 02139, USA
\item[\panjab] Panjab University, Chandigarh 160 014, India.
\item[\debrecen] KLTE-ATOMKI, H-4010 Debrecen, Hungary$^\P$
\item[\dublin] Department of Experimental Physics,
  University College Dublin, Belfield, Dublin 4, Ireland
\item[\florence] INFN Sezione di Firenze and University of Florence, 
     I-50125 Florence, Italy
\item[\cern] European Laboratory for Particle Physics, CERN, 
     CH-1211 Geneva 23, Switzerland
\item[\wl] World Laboratory, FBLJA  Project, CH-1211 Geneva 23, Switzerland
\item[\geneva] University of Geneva, CH-1211 Geneva 4, Switzerland
\item[\hefei] Chinese University of Science and Technology, USTC,
      Hefei, Anhui 230 029, China$^{\triangle}$
\item[\lausanne] University of Lausanne, CH-1015 Lausanne, Switzerland
\item[\lyon] Institut de Physique Nucl\'eaire de Lyon, 
     IN2P3-CNRS,Universit\'e Claude Bernard, 
     F-69622 Villeurbanne, France
\item[\madrid] Centro de Investigaciones Energ{\'e}ticas, 
     Medioambientales y Tecnol\'ogicas, CIEMAT, E-28040 Madrid,
     Spain${\flat}$ 
\item[\florida] Florida Institute of Technology, Melbourne, FL 32901, USA
\item[\milan] INFN-Sezione di Milano, I-20133 Milan, Italy
\item[\moscow] Institute of Theoretical and Experimental Physics, ITEP, 
     Moscow, Russia
\item[\naples] INFN-Sezione di Napoli and University of Naples, 
     I-80125 Naples, Italy
\item[\cyprus] Department of Physics, University of Cyprus,
     Nicosia, Cyprus
\item[\nymegen] University of Nijmegen and NIKHEF, 
     NL-6525 ED Nijmegen, The Netherlands
\item[\caltech] California Institute of Technology, Pasadena, CA 91125, USA
\item[\perugia] INFN-Sezione di Perugia and Universit\`a Degli 
     Studi di Perugia, I-06100 Perugia, Italy   
\item[\peters] Nuclear Physics Institute, St. Petersburg, Russia
\item[\cmu] Carnegie Mellon University, Pittsburgh, PA 15213, USA
\item[\potenza] INFN-Sezione di Napoli and University of Potenza, 
     I-85100 Potenza, Italy
\item[\prince] Princeton University, Princeton, NJ 08544, USA
\item[\riverside] University of Californa, Riverside, CA 92521, USA
\item[\rome] INFN-Sezione di Roma and University of Rome, ``La Sapienza",
     I-00185 Rome, Italy
\item[\salerno] University and INFN, Salerno, I-84100 Salerno, Italy
\item[\ucsd] University of California, San Diego, CA 92093, USA
\item[\sofia] Bulgarian Academy of Sciences, Central Lab.~of 
     Mechatronics and Instrumentation, BU-1113 Sofia, Bulgaria
\item[\korea]  The Center for High Energy Physics, 
     Kyungpook National University, 702-701 Taegu, Republic of Korea
\item[\purdue] Purdue University, West Lafayette, IN 47907, USA
\item[\psinst] Paul Scherrer Institut, PSI, CH-5232 Villigen, Switzerland
\item[\zeuthen] DESY, D-15738 Zeuthen, Germany
\item[\eth] Eidgen\"ossische Technische Hochschule, ETH Z\"urich,
     CH-8093 Z\"urich, Switzerland
\item[\hamburg] University of Hamburg, D-22761 Hamburg, Germany
\item[\taiwan] National Central University, Chung-Li, Taiwan, China
\item[\tsinghua] Department of Physics, National Tsing Hua University,
      Taiwan, China
\item[\S]  Supported by the German Bundesministerium 
        f\"ur Bildung, Wissenschaft, Forschung und Technologie
\item[\ddag] Supported by the Hungarian OTKA fund under contract
numbers T019181, F023259 and T037350.
\item[\P] Also supported by the Hungarian OTKA fund under contract
  number T026178.
\item[$\flat$] Supported also by the Comisi\'on Interministerial de Ciencia y 
        Tecnolog{\'\i}a.
\item[$\sharp$] Also supported by CONICET and Universidad Nacional de La Plata,
        CC 67, 1900 La Plata, Argentina.
\item[$\triangle$] Supported by the National Natural Science
  Foundation of China.
\end{list}
}
\vfill


\newpage

\end{document}